\documentclass[usenatbib]{mnras}

\usepackage{newtxtext,newtxmath}

\usepackage[T1]{fontenc}
\usepackage{ae,aecompl}

\usepackage{graphicx}	
\usepackage{amsmath}	
\usepackage{amssymb}	

\usepackage{siunitx}
\usepackage{comment}

\PassOptionsToPackage{draft}{hyperref}
\usepackage{xspace}

\usepackage[xindy, toc, hyperfirst=false, nolist, nostyles, sanitize={name=false,description=false,symbol=false}]{glossaries}
\glsdisablehyper
\usepackage[hyperref,x11names, table]{xcolor}

\newcommand{\angstrom}{\mbox{\normalfont\AA}}

\newcommand{\msun}{\hbox{$M_{\odot}$}}

\newglossaryentry{vrad}{name={radial velocity~}, text={radial velocity}, symbol={\ensuremath{v_\textrm{rad}}}, description={radial velocity}, sort=vrad}
\newglossaryentry{vrot}{name={stellar rotation~}, name={stellar rotation}, symbol={\ensuremath{v_\textrm{rot}}}, description={radial velocity}, sort=vrot}
\newcommand{\vrot}{\glssymbol*{vrot}}
\newcommand{\vrad}{\glssymbol*{vrad}}

\newcommand{\kms}{\ensuremath{\textrm{km}~\textrm{s}^{-1}}}

\newcommand{\xray}{X-ray}

\newglossaryentry{angstrom}{name=\AA, description={unit of length $10^{-10}$\,m}, sort=angstrom}
\newglossaryentry{nir}{name=NIR,description={near infrared},first = {near infrared (NIR)}}
\newglossaryentry{psf}{name=PSF,description={point-spread function},first = {point-spread function (PSF)}}
\newglossaryentry{fwhm}{name=FWHM,description={Full Width Half Maximum},first = {FWHM}}
\newglossaryentry{rms}{name=RMS,description={Root Mean Square},first = {RMS}}
\newglossaryentry{signalnoise}{name=S/N,description={signal to noise}}
\newglossaryentry{uv}{name=UV,description={ultra violet},first = {ultra violet (UV)}}
\newglossaryentry{halpha}{name=\ensuremath{\textrm{H}\alpha}, description={First line of the Balmer series at 6563\,\AA}, sort=halpha}
\newglossaryentry{mgb}{name={Mg \textsc{i} b}, description={Triplet at 5167\,\AA, 5173\,\AA and 5184\,\AA}}
\newglossaryentry{sobolevapprox}{name={Sobolev approximation}, description={Lines are approximation with an infinitley thin interaction region \citep[e.g. no broadening][]{1960mes..book.....S}}, first={Sobolev approximation }}
\newglossaryentry{radeq}{name={radiative equilibrium}, description={The net flux of energy between matter and radiation field is zero}}
\newglossaryentry{nebularapprox}{name={nebular approximation}, description={Assumes that the plasma condition are controlled by a central radiation source. The radiation field decreases with the distance to the source by geometrical dilution. See \citet{1978stat.book.....M} for details}}
\newglossaryentry{modnebularapprox}{name={modified nebular approximation}, description={In contrast to \gls{nebularapprox} where only geometrical dilution is taken into account, the modified nebular approximation also takes dilution by other radiative processes into account }, first={modified nebular approximation}, parent=nebularapprox}
\newglossaryentry{thompsonscat}{name={Thomson scattering}, description={Scattering of photons on low energy electrons}}
\newglossaryentry{lte}{name={LTE}, description={Local Thermodynamic Equilibrium}, first={local thermodynamic equilibrium (LTE)}}
\newglossaryentry{lsr}{name={LSR}, description={Local Standard of Rest}, first={\textit{local standard of rest} (LSR)}}
\newglossaryentry{mc}{name={MC}, description={Monte Carlo}, first={\textit{Monte Carlo} (MC)}}
\newglossaryentry{wcs}{name={WCS}, description={world coordinate system}, first={world coordinate system (WCS)}}
\newglossaryentry{cmf}{name=CMF, text=CMF, first=Comoving Frame (CMF henceforth), description={Comoving Frame}}

\newglossaryentry{uvoir}{name=UVOIR, text=UVOIR, first=UV/optical/Near-IR (UVOIR), description={UV/optical/Near-IR}}

\newglossaryentry{sfit}{name=SFIT, text=\textsc{sfit}, description={spectral fitting program for hot stars \citep{2001A&A...376..497J}}, first={\textsc{sfit} \citep{2001A&A...376..497J}}}
\newglossaryentry{iraf}{name=IRAF, text=\textsc{iraf}, description={Image Reduction and Analysis Facility maintained by NOAO}, first={\textsc{iraf}\protect\footnote{IRAF: the Image Reduction and Analysis Facility is distributed by the National Optical Astronomy Observatory, which is operated by the Association of Universities for Research in Astronomy (AURA) under cooperative agreement with the National Science Foundation (NSF).}}}
\newglossaryentry{pyraf}{name=PyRAF, text=\textsc{PyRAF}, description={Python wrap of \gls{iraf} maintained by STSCI}, first=\textsc{PyRAF} \protect\footnote{PyRAF is a product of the Space Telescope Science Institute, which is operated by AURA for NASA.}}
\newglossaryentry{astropy}{name=ASTROPY, text=\textsc{astropy}, description=\textsc{astropy} framework, first = \textsc{astropy} \citep{2013A&A...558A..33A}}
\newglossaryentry{numpy}{name=NUMPY, text=\textsc{numpy}, description=\textsc{numpy} framework, first = \textsc{numpy} \citep{walt2011numpy}}
\newglossaryentry{scipy}{name=SCIPY, text=\textsc{scipy}, description=\textsc{scipy} framework, first = \textsc{scipy} \citep{Jones:2001fk}}
\newglossaryentry{matplotlib}{name=matplotlib, text=\textsc{matplotlib}, description=\textsc{matplotlib} framework, first = \textsc{matplotlib} \citep{hunter2007matplotlib}}
\newglossaryentry{pandas}{name=pandas, text=\textsc{pandas}, description=\textsc{pandas} framework, first = \textsc{pandas} \citep{mckinney2010data}}
\newglossaryentry{ipython}{name=ipython, text=\textsc{ipython}, description=\textsc{ipython} framework, first = \textsc{ipython} \citep{perez2007ipython}}
\newglossaryentry{jupyter}{name=jupyter, text=\textsc{jupyter}, description=\textsc{jupyter} framework, first = \textsc{jupyter} \citep{kluyver2016jupyter,perez2015project,ragan2014jupyter}}
\newglossaryentry{aplpy}{name=aplpy, text=\textsc{aplpy}, description=\textsc{aplpy} framework, first = \textsc{aplpy} \citep{2012ascl.soft08017R}}
\newglossaryentry{nltk}{name=nltk, text=\textsc{nltk}, description=\textsc{nltk} framework, first = Natural Language ToolKit \citep[\textsc{NLTK};][]{bird2009natural}}
\newglossaryentry{scikit-learn}{name=scikit-learn, text=\textsc{scikit-learn}, description=\textsc{scikit-learn} framework, first = \textsc{scikit-learn} \citep[][]{scikit-learn}}
\newglossaryentry{scikit-image}{name=scikit-image, text=\textsc{scikit-image}, description=\textsc{scikit-image} framework, first = \textsc{scikit-image} \citep[][]{scikit-image}}
\newglossaryentry{moog}{name=MOOG,text={\textsc{moog}}, description={spectral synthesis software \citep{1973ApJ...184..839S}}, first={\textsc{Moog} \citep{1973ApJ...184..839S}}}
\newglossaryentry{atlas9}{name=ATLAS9,description={grid of stellar atmospheres \citep{2004astro.ph..5087C}}, first={ATLAS9 \citep{2004astro.ph..5087C}}}
\newglossaryentry{vald}{name=VALD,description={Vienna Atomic Line Database \citep{2000BaltA...9..590K}}, first={Vienna Atomic Line Database \citep[VALD;][]{2000BaltA...9..590K}}}
\newglossaryentry{sextractor}{name=SExtractor, text=\textsc{SExtractor}, description={Source Extractor photometry program \citep{1996A&AS..117..393B}}, first={\textsc{SExtractor} \citep{1996A&AS..117..393B}}}

\newglossaryentry{swarp}{name=SWarp, text=\textsc{SWarp}, description={SWarp \citep{2002ASPC..281..228B}}, first={\textsc{SWarp} \citep{2002ASPC..281..228B}}}
\newglossaryentry{astrometry.net}{name=astrometry.net, text=\textsc{astrometry.net}, description={\textsc{astrometry.net} \citep{2010AJ....139.1782L}} first={\textsc{astrometry.net} \citep{2010AJ....139.1782L}}}

\newglossaryentry{astrodrizzle}{name=AstroDrizzle, text=\textsc{AstroDrizzle}, description={AstroDrizzle \citep{2012drzp.book.....G}}, first={\textsc{AstroDrizzle} \citep{2012drzp.book.....G}}}

\newglossaryentry{idl}{name=IDL,text={\textsc{idl}}, description={Interactive Data Language}}
\newglossaryentry{makee}{name=MAKEE,text=\textsc{makee}, description={MAuna Kea Echelle Extraction by Tom Barlow available}}
\newglossaryentry{minuit}{name=MINUIT,text={\textsc{minuit}}, description={collection of numerical optimization tools \citep{James:1975dr}}}
\newglossaryentry{migrad}{name=MIGRAD,text={\textsc{migrad}}, description={numerical gradient optimization tools - part of \gls{minuit}}}
\newglossaryentry{dolphot}{name=DOLPHOT, text=\textsc{dolphot}, description=photometry package for HST, first=\textsc{dolphot} \citep{2000PASP..112.1383D}}
\newglossaryentry{synphot}{name=synphot, text={\textsc{synphot}}, description={synthetic photometry package from STSCI}, first={\textsc{synphot}\protect\footnote{\textsc{synphot} is a product of the Space Telescope Science Institute, which is operated by AURA for NASA.}}}
\newglossaryentry{chianti}{name=CHIANTI, text=CHIANTI, description= CHIANTI Database 7.1, first =CHIANTI 7.1 \citep{1997A&AS..125..149D,2012ApJ...744...99L}}
\newglossaryentry{synpp}{name=SYNPP, text=SYN++, description= SYN++ software, first =SYN++ \citep{2011PASP..123..237T}}
\newglossaryentry{tardis}{name=TARDIS, text=\textsc{tardis}, description= TARDIS MC code, first = {\textsc{tardis} \citep{2014MNRAS.440..387K}}}

\newglossaryentry{artis}{name=ARTIS, text=\textsc{artis}, description= ARTIS MC code, first = \textsc{artis} \citep{2009MNRAS.398.1809K}}
\newglossaryentry{cmfgen}{name=CMFGEN, text=\textsc{cmfgen}, description=CMFGGEn radiative transfer code, first = \textsc{cmfgen} \citep{1998ApJ...496..407H}}
\newglossaryentry{sedona}{name=SEDONA, text=\textsc{sedona}, description= Sedona MC code, first = \textsc{sedona} \citep{2006ApJ...651..366K}}
\newglossaryentry{mlmc}{name=MLMC, text=ML93, description= Mazzali Lucy Monte Carlo, first ={Mazzali \& Lucy (1993, ML93) code}}
\newglossaryentry{starkit}{name=STARKIT, text=\textsc{starkit}, description= TARDIS MC code, first = {\textsc{starkit} \citep{wolfgang_kerzendorf_2015_28016}}}

\newglossaryentry{pyne}{name=PYNE, text=\textsc{pyne}, description= PYNE code, first = {\textsc{pyne} \citep{Scopatz2012a}}}
\newglossaryentry{multinest}{name=MULTINEST, text=\textsc{MultiNest}, description=MultiNest, first={\textsc{MultiNest} \citep{2009MNRAS.398.1601F}}}
\newglossaryentry{wsynphot}{name=WSYNPHOT, text=\textsc{wsynphot}, description=Wsynphot, first={\textsc{wsynphot}\protect\footnote{\protect\url{https://github.com/wkerzendorf/wsynphot}}}}
\newglossaryentry{specutils}{name=SPECUTILS, text=\textsc{specutils}, description=specutils, first={\textsc{specutils} \protect\footnote{\protect\url{https://github.com/astropy/specutils}}}}
\newglossaryentry{ads}{name=ADS ,description=ADS, first={NASA Astrophysics Data System (ADS) \citep{2000A&AS..143...41K}}}

\newglossaryentry{2mass}{name=2MASS,description={Two Micron All Sky Survey \citep{2006AJ....131.1163S}}, first={Two Micron All Sky Survey \citep{2006AJ....131.1163S}}}
\newglossaryentry{nomad}{name=NOMAD,first={Naval Observatory Merged Astrometric Dataset \citep[NOMAD; ][]{2005yCat.1297....0Z}}, description={Naval Observatory Merged Astrometric Dataset}}
\newglossaryentry{wifes}{name=WIFES, text=\textsc{WiFeS}, first={\textsc{WiFeS} \citep{2007Ap&SS.310..255D}},  description={Wide Field Spectrograph - \gls{ifu} mounted on the 2.3\,m telescope at Siding Spring Observatory}}
\newglossaryentry{scp}{name=SCP,description={Supernova Cosmology Project, led by Saul Perlmutter}, first={Supernova Cosmology Project (SCP)}}
\newglossaryentry{hzsns}{name=HZSNS,description={High Z Supernova Search, led by Brian Schmidt}, first={High Z Supernova Search (HZSNS)}}
\newglossaryentry{vlt}{name=VLT,description={Very Large Telescope located on Cerro Paranal (Chile)}, first={Very Large Telescope (VLT)}}
\newglossaryentry{flames}{name=FLAMES,description={Multi-object, intermediate and high resolution spectrograph mounted on the  \gls{vlt}}}
\newglossaryentry{hires}{name=HIRES, description={High Resolution Echelle Spectrometer mounted on the Keck Telescope}, first={High Resolution Echelle Spectrometer \citep[HIRES;][]{1994SPIE.2198..362V}}}
\newglossaryentry{lris}{name=LRIS,description={Low Resolution Imaging Spectrometer mounted on the Keck Telescope}, first={Low-Resolution Imaging Spectrometer \citep[LRIS;][]{Oke95}}}

\newglossaryentry{decam}{name=DECam, description={DECam is a high-performance, wide-field CCD imager mounted at the prime focus of the Blanco 4-m telescope at \gls{ctio}.}, first={Dark Energy Camera \citep[DECam; ][]{2012PhPro..37.1332D,2015AJ....150..150F}}}

\newglossaryentry{essence}{name=ESSENCE,description={The `Equation of State: SupErNovae trace Cosmic Expansion' project \citep[ESSENCE;][]{2002AAS...201.7809G}}, first={`The Equation of State: SupErNovae trace Cosmic Expansion' \citep[ESSENCE;][]{2002AAS...201.7809G}}}
\newglossaryentry{ifu}{name=IFU,description={Optical instrument combining spectrographic and imaging capabilities, used to obtain spatially resolved spectra}, first={Integral Field Unit (IFU)}, firstplural={Integral Field Units (IFUs)}}

\newglossaryentry{besancon}{name=Besan\c{c}on Model, description={Model of stellar population synthesis of the Galaxy, including kinematics.}}

\newglossaryentry{int}{name=INT,description={Isaac Newton 2.5\,m Telescope}, first={Isaac Newton 2.5\,m Telescope (INT)}}
\newglossaryentry{iau}{name=IAU,description={International Astronomical Union}, first={IAU}}
\newglossaryentry{chandra}{name=Chandra,description={Chandra \xray\ Observatory (space-based)}}
\newglossaryentry{hst}{name=HST,description={Hubble Space Telescope}}
\newglossaryentry{hst.wfpc2}{name=WFPC2,description={Wide-Field Planetary Camera 2 mounted on the \gls{hst}}, first={Wide-Field Planetary Camera 2 (WFPC2)}}
\newglossaryentry{hst.acs}{name=ACS,description={Advanced Camera for Surveys mounted on the \gls{hst}}, first={Advanced Camera for Surveys (ACS)}}
\newglossaryentry{hst.wfc3}{name=WFC3,description={Wide-Field Camera 3 mounted on the \gls{hst}}, first={Wide-Field Camera 3 (WFC3)}}
\newglossaryentry{hst.cte}{name=CTE, description={charge transfer efficiency (CTE)}, first={charge transfer efficiency \citep[CTE; see ][for a description]{2009acs..rept....1C}}}

\newglossaryentry{snls}{name=SNLS,description={Supernova Legacy Survey \citep{2003AAS...203.8209P}}, first={Supernova Legacy Survey \citep[SNLS;][]{2003AAS...203.8209P}}}
\newglossaryentry{dass}{name=DASS, description={Digitized Astronomy Supernova Survey \citep{1975PASP...87..565C}}, first={Digitized Astronomy Supernova Survey \citep[DASS;][]{1975PASP...87..565C}}}
\newglossaryentry{bait}{name=BAIT, description={Berkley Automatic Imaging Telescope \citep{1993PASP..105.1164R}}, first={Berkley Automatic Imaging Telescope \citep[BAIT;][]{1993PASP..105.1164R}}}
\newglossaryentry{kait}{name=KAIT, description={Katzman Automatic Imaging Telescope \citep{2001ASPC..246..121F}}, first={Katzman Automatic Imaging Telescope \citep[KAIT;][]{2001ASPC..246..121F}}}
\newglossaryentry{loss}{name=LOSS, description={Lick Observatory Supernova Search  \citep{2000AIPC..522..103L}}, first={Lick Observatory Supernova Search \citep[LOSS;][]{2000AIPC..522..103L}}}
\newglossaryentry{ctss}{name=CTSS,description={Cal\'{a}n/Tololo Supernova Survey \citep{1993AJ....106.2392H}}, first={Cal\'{a}n/Tololo supernova survey \citep[CTSS;][]{1993AJ....106.2392H}}}
\newglossaryentry{ctio}{name= CTIO, description={Cerro Tololo Inter-American Observatory}, first={Cerro Tololo Inter-American Observatory (CTIO)}}
\newglossaryentry{ptf}{name=PTF, description={Palomar Transient Factory \citep{2009PASP..121.1334R}}, first={Palomar Transient Factory \citep[PTF;][]{2009PASP..121.1334R}}}
\newglossaryentry{batse}{name=BATSE, description={Burst and Transient Source Experiment mounted on the Compton Gamma Ray Observatory}, first={Burst and Transient Source Experiment (BATSE)}}
\newglossaryentry{bepposax}{name=BeppoSAX, description={\xray\ satellite named in honor of Giuseppe "Beppo" Occhialini}}
\newglossaryentry{rosat}{name=ROSAT, description={short for R\"{o}ntgensatellit}, first={ROSAT}}
\newglossaryentry{hete2}{name=HETE2, description={High Energy Transient Explorer}, first={High Energy Transient Explorer (HETE)}}
\newglossaryentry{ska}{name=SKA, description={Square Kilometre Array}, first={Square Kilometre Array (SKA)}}

\newglossaryentry{gnirs}{name=GNIRS, description={Gemini Near InfraRed Spectrograph mounted on the Gemini North Telescope}}
\newglossaryentry{gmosn}{name=GMOS, description={Gemini Multi Object Spectrograph mounted on the
 Gemini North Telescope}, first={GMOS \citep[Gemini Multi Object Spectrograph;][]{2004PASP..116..425H}}}
\newglossaryentry{swift}{name=Swift, description={Swift Gamma-Ray Burst Mission}}
\newglossaryentry{vla}{name=VLA, description={Very Large Array radio telescope located in North America}, first={Very Large Array (VLA)}}
\newglossaryentry{evla}{name=EVLA, description={Extended Very Large Array radio telescope located in North America}, first={Extended Very Large Array (EVLA)}}
\newglossaryentry{sdss}{name=SDSS, description={Sloan Digital Sky Survey}}
\newglossaryentry{dss}{name=DSS, description={Digitized Sky Survey}}
\newglossaryentry{skymapper}{name=SkyMapper, description={SkyMapper telescope \citep{2007PASA...24....1K}}, first={SkyMapper \citep{2007PASA...24....1K}}}
\newglossaryentry{panstarrs}{name=PanSTARRS, description={Panoramic Survey Telescope \& Rapid Response System \citep{2004SPIE.5489...11K}}, first={Panoramic Survey Telescope \& Rapid Response System \citep[PanSTARRS;][]{2004SPIE.5489...11K}}}
\newglossaryentry{ps1dr1}{name=PS1~DR1, description={Panoramic Survey Telescope \& Rapid Response System \citep{2004SPIE.5489...11K} }, first={Panoramic Survey Telescope \& Rapid Response System \citep[PanSTARRS;][]{2004SPIE.5489...11K} DR1}}

\newglossaryentry{lsst}{name=LSST, description={Large Synoptic Survey Telescope}, first={Large Synoptic Survey Telescope \citep[LSST;][]{2006AAS...209.8604P}}}
\newglossaryentry{ppmxl}{name=PPMXL, description={PPMXL Catalog of Positions and Proper Motions on the ICRS \citep{2010AJ....139.2440R}}}
\newglossaryentry{gaia}{name=GAIA, description={Global Astrometric Interferometer for Astrophysics \citep{2001A&A...369..339P}}, first={Global Astrometric Interferometer for Astrophysics \citep[GAIA;][]{2001A&A...369..339P}}}
\newglossaryentry{ligo}{name=LIGO, description={Laser Interferometer Gravitational Wave Observatory}, first={Laser Interferometer Gravitational Wave Observatory \citep[LIGO;][]{1992Sci...256..325A}}}
\newglossaryentry{aligo}{name=Advanced LIGO, description={Advanced LIGO}, sort=ligo2}
\newglossaryentry{lisa}{name=LISA, description={Laser Interferometer Space Antenna \citep{1994ESAJ...18..219J}}, first={Laser Interferometer Space Antenna \citep[LISA;][]{1994ESAJ...18..219J}}}
\newglossaryentry{ccd}{name=CCD,description={Charged Coupled Device}, first={charged coupled device (CCD)}, firstplural={charged coupled devices (CCDs)}}

\newcommand{\sn}[2]{SN~#1#2\xspace}

\newglossaryentry{irc}{name=IRC, text={IRC}, description={infrared catastrophe}, first={infrared catastrophe \citep[IRC;][]{1980PhDT.........1A}}}

\newglossaryentry{sn}{name=Supernova, text={SN}, plural={SNe}, description={exploding star}, nonumberlist=true, first={supernova (SN)}, firstplural={supernovae (SNe)}}
\newglossaryentry{snia}{name=Type~Ia (SN~Ia), text={SN~Ia}, description={Thermonuclear explosion of a white dwarf - spectra show no hydrogen but a strong silicon line},first={Type~Ia supernova (SN~Ia)}, firstplural={Type Ia supernovae (SNe~Ia)}, plural={SNe~Ia}, parent=sn, nonumberlist=true}
\newcommand{\sneia}{\glspl*{snia}\xspace}
\newcommand{\snia}{\gls*{snia}\xspace}

\newglossaryentry{branchnormal}{name={branch-normal}, text=\textit{Branch-normal}, description={Large homogeneous class of Type Ia Supernovae, defined in \citet{1993AJ....106.2383B}}, first={\textit{Branch-normal} SNe Ia \citep{1993AJ....106.2383B}}, parent=snia}
\newglossaryentry{91t}{name={91T-like}, description={Luminous class of Type Ia supernovae similar to \sn{1991}{T} \citep{1992AJ....103.1632P}} , first={91T-like}, parent=snia}
\newglossaryentry{91bg}{name={91bg-like}, description={Faint class of Type Ia supernovae similar to \sn{1991}{bg} \citep{1992AJ....104.1543F}}, first={91bg-like}, parent=snia}
\newglossaryentry{02cx}{name={02cx-like}, description={Peculiar class of Type Ia supernovae similar to \sn{2002}{cx} \citep{2003PASP..115..453L}}, first={02cx-like \sneia\ \citep{2003PASP..115..453L}}, parent=snia}

\newglossaryentry{snibc}{name=Type~Ib/c, text={SN~Ib/c}, description={Collapse of the core of a massive star -  spectrum shows no hydrogen and no silicon line},first={Type~Ib/c supernova (SN~Ib/c)}, firstplural={Type~Ib/c supernovae (SNe~Ib/c)}, plural={SNe~Ib/c}, parent=sn}

\newglossaryentry{snib}{name=Type~Ib, text={SN~Ib}, description={Spectrum shows no hydrogen and no silicon, but helium line},first={Type Ib supernova (SN~Ib)}, firstplural={Type~Ib supernovae (SNe~Ib)}, plural={SNe~Ib}, parent=snibc}

\newglossaryentry{snic}{name=Type~Ic, text={SN~Ic}, description={Spectrum shows no hydrogen, no silicon and no helium line},first={Type~Ic supernova (SN~Ic)}, firstplural={Type~Ic supernovae (SNe~Ic)}, plural={SNe~Ic}, parent=snibc}

\newglossaryentry{snii}{name=Type~II, text={SN~II}, description={Collapse of the core of a massive star - spectrum shows strong hydrogen line},first={Type~II supernova (SN~II)}, firstplural={Type~II supernovae (SNe~II)}, plural={SNe~II}, parent=sn}

\newglossaryentry{sniib}{name=Type~IIb, text={SN~IIb}, description={Spectrum shows hydrogen and helium lines},first={Type~IIb supernova (SN~IIb)}, firstplural={Type~IIb supernovae (SNe~IIb)}, plural={SNe~IIb}, parent=snii}

\newglossaryentry{sniip}{name=Type~II~Plateau (Type IIP), text={SN~IIP}, description={Lightcurve shows plateau},first={Type~IIP supernova (SN~IIP)}, firstplural={Type~II Plateau supernovae \citep[SNe~IIP;][]{1979A&A....72..287B}}, plural={SNe~IIP}, parent=snii}

\newglossaryentry{sniil}{name=SN~II~Linear, text={SN~IIL}, description={Lightcurve shows no plateau, but linear decline},first={Type~IIL supernova (SN~IIL)}, firstplural={Type~II~Linear supernovae \citep[SNe~IIL;][]{1990MNRAS.244..269S}}, plural={SNe~IIL}, parent=snii}

\newglossaryentry{sniin}{name=Type II narrow-lined (Type IIn), description={Spectrum shows narrow lines},first={Type~II~narrow-lined supernova (SN IIn)}, firstplural={Type~IIn supernovae (SNe~IIn)}, plural={SNe~IIn}, parent=snii}

\newglossaryentry{snr}{name=Remnant (SNR), text=SNR, description={Remnant left visible post-explosion}, first={supernova remnant (SNR)}, firstplural={supernova remnants (SNRs)}, parent=sn}

\newglossaryentry{dtd}{name=DTD,description={delay time distribution - expected supernova rate over time after a brief outburst of starformation},first={delay time distribution (DTD)}, firstplural={delay time distributions (DTDs)}, plural=DTDs}

\newglossaryentry{hvg}{name=HVG,description={high velocity gradient - Type Ia supernovae with a fast evolution of photospheric velocity},first={high velocity group (HVG)}, firstplural={high velocity groups (HVGs)}, plural=HVGs, parent=snia}

\newglossaryentry{lvg}{name=LVG,description={low velocity gradient - Type Ia supernovae with a slow evolution of photospheric velocity},first={low velocity group (LVG)}, firstplural={low velocity groups (LVGs)}, plural=LVGs, parent=snia}

\newglossaryentry{wd}{name=white dwarf (WD), text=WD, description={White Dwarf - extremely dense stellar remnant}, first={white dwarf (WD)}}
\newglossaryentry{onemgwd}{name= Oxygen/Neon (ONe), text={ONe-WD},description={Oxygen/Neon White Dwarf}, first={oxygen/neon White Dwarf (ONe-WD)}, parent=wd}
\newglossaryentry{cowd}{name=carbon/oxygen (CO), text={CO-WD}, description={carbon/oxygen white dwarf}, first={carbon/oxygen white dwarf (CO-WD)}, firstplural = {carbon/oxygen white dwarfs (CO-WDs)}, parent=wd}

\newglossaryentry{sds}{name=SD-Scenario,description={single-degenerate scenario (single white dwarf accreting from non-degenerate companion)}, first={single-degenerate scenario (SD-scenario)}}

\newglossaryentry{dds}{name=DD-Scenario, description={double degenerate scenario (merging of two white dwarfs)}, first={double-degenerate scenario (DD-scenario)}}

\newglossaryentry{sss}{name=SSS, text={supersoft \xray\ source}, description={supersoft \xray\ source - believed to be emitted by nuclear fusion on a white dwarf's surface}}

\newglossaryentry{amcvn}{name=AM CVn, description={AM Canum Venaticorum star \citep[white dwarf accreting hydrogen poor matter from a companion star; see ][]{2005ASPC..330...27N}}}

\newglossaryentry{rlof}{name=RLOF, description={Roche Lobe Overflow (see \citet{1971ARA&A...9..183P} for a more detailed description)}, first={Roche-lobe overflow (RLOF)}}

\newglossaryentry{mchan}{name={Chandrasekhar mass~}, text={Chandrasekhar~mass}, symbol={\ensuremath{M_\textrm{Chan}}}, plural={Chandrasekhar~masses}, description={Mass when the core of a star collapses due to insufficient degeneracy pressure - for a white dwarf $\approx1.38\,M_\odot$ see \citet{1931ApJ....74...81C}}, first={Chandrasekhar~mass \citep[$M_\textrm{Chan}=1.38\,M_\odot$;][]{1931ApJ....74...81C}}, sort=mchan}

\newglossaryentry{w7}{name={W7 model},description={W7 model \citep{1984ApJ...286..644N}},first = {W7 model \citep{1984ApJ...286..644N}}}

\newglossaryentry{ew}{name=Equivalent Width, text={EW}, description={width of a rectangle that has the same area as a spectral line when taken to zero flux}, first={equivalent width (EW)}, firstplural={equivalent widths (EWs)}}
\newglossaryentry{agb}{name=AGB,description={Asymptotic Giant Branch}, first={Asymptotic Giant Branch (AGB)}}
\newglossaryentry{cmb}{name=CMB,description={Cosmic Microwave Background}}
\newglossaryentry{csm}{name=CSM,description={Circumstellar Medium}, first={circumstellar medium (CSM)}}
\newglossaryentry{csi}{name=CSI,description={Circumstellar Interaction}, first={circumstellar interaction (CSI)}}
\newglossaryentry{ism}{name=ISM,description={Interstellar Medium}, first={interstellar medium (ISM)}}
\newglossaryentry{ige}{name=IGE,description={Iron Group Element}, first={iron group element (IGE)}, firstplural={iron group elements (IGEs)}}
\newglossaryentry{epm}{name=EPM,description={Expanding Photosphere Method \citep{1974ApJ...193...27K}}, first={Expanding Photosphere Method (EPM)}}
\newglossaryentry{aic}{name=AIC,description={Accretion Induced Collapse}, first={accretion induced collapse (AIC)}}
\newglossaryentry{ime}{name=IME,description={Intermediate Mass Element}, first={intermediate mass element (IME)}, firstplural={intermediate mass elements (IMEs)}}
\newglossaryentry{h0}{name=\ensuremath{H_0},description={Hubbles constant}}
\newglossaryentry{nse}{name=NSE,description={Nuclear Statistical Equilibrium}, first={nuclear statistical equilibrium (NSE)}}
\newglossaryentry{cdm}{name=CDM,description={Cold Dark Matter}}
\newglossaryentry{grb}{name=GRB,description={Gamma Ray Burst}, first={Gamma Ray Burst (GRB)}, firstplural={Gamma Ray Bursts (GRBs)}}
\newglossaryentry{xps}{name=XPS, description={x-ray point source}, first={x-ray point source (XPS)}, firstplural={x-ray point sources (XPS)}}
\newglossaryentry{donor}{name=donor,description={non-degenerate companion in the \gls{sds}}}
\newglossaryentry{mainsequence}{name=main sequence,description={main sequence star}}
\newglossaryentry{redgiant}{name=red giant,description={red giant star}}
\newglossaryentry{mlcs}{name=MLCS,description={Multicolor Light Curve Shape method \citep[MLCS;][]{1996ApJ...473...88R}}, first={Multicolor Light-Curve Shape method \citep[MLCS;][]{1996ApJ...473...88R}}}
\newglossaryentry{rsoph}{name=RS~Ophiuci ,description={white dwarf accreting from a red giant - assumed progenitor of the \gls{sds}}, sort=rsoph}
\newglossaryentry{usco}{name=U~Scorpii,description={white dwarf accreting from a main sequence star - assumed progenitor of the \gls{sds}}, sort=usco}
\newglossaryentry{rcw86}{name=RCW~86,description={supernova remnant sometimes associated with \sn{185}{}}, sort=rcw86}
\newglossaryentry{casa}{name=Cas~A,description={Cassiopeia A supernova remnant - probably a \gls{snib} event}}
\newglossaryentry{cepheid}{name=Cepheid,description={very luminous variable star with a strong luminosity period relationship}}
\newglossaryentry{urca}{name=Urca, text=\textit{Urca}, description={process predominatly contributing to cooling in stars. The \textit{Urca} process consists of alternating electron-capture and $\beta^{-}$ decay of two nuclei pairs.},sort=urca}
\newglossaryentry{alphacen}{name=Alpha Centauri,description={one of the brightest stars in the night sky and a close binary}}
\newglossaryentry{pcygni}{name={P Cygni}, text={P Cygni},description={a hypergiant luminous blue variable with strong winds. Often referred to as a description for their line profiles showing a emission peak at the rest wavelength of the line and a blue-shifted absorption trough.}}

\newglossaryentry{teff}{name={effective temperature~}, text={effective temperature}, symbol={\ensuremath{T_\textrm{eff}}}, description={Temperature of a blackbody emitting the same total energy}, sort=teff}

\newglossaryentry{logg}{name={surface gravity~}, text={surface gravity}, symbol={\ensuremath{\textrm{log}\,g}}, description={gravity at the surface of a star}, sort=logg}
\newglossaryentry{feh}{name={metallicity~}, text={metallicity}, symbol=\textrm{[Fe/H]},description={iron abundance relative to the sun}, sort=feh}

\newglossaryentry{texp}{name={time since explosion~}, text={time since explosion}, text={time since explosion}, symbol={\ensuremath{t_{\rm exp}}},description={time since explosion (measured in days)}, sort=texp, first={time since explosion (\ensuremath{t_{\rm exp}})}}

\newglossaryentry{lmc}{name=LMC,description={Large Magellanic Cloud}, first={Large Magellanic Cloud (LMC)}, sort=lmc}
\newglossaryentry{smc}{name=SMC,description={Small Magellanic Cloud}, sort=smc}
\newglossaryentry{z}{name=\ensuremath{z},description={redshift}, sort=z}
\newcommand{\teff}{\glssymbol*{teff}}
\newcommand{\logg}{\glssymbol*{logg}}
\newcommand{\feh}{\glssymbol*{feh}}

\newglossaryentry{stats.pdf}{name=PDF, description={Probability Density Function}, first={Probability Density Function}}

\title[Tycho-B: an unlikely companion for SN 1572]{Tycho-B: an unlikely companion for SN 1572\thanks{Based on observations made with the NASA/ESA Hubble Space Telescope, obtained at the Space Telescope Science Institute, which is operated by the Association of Universities for Research in Astronomy, Inc., under NASA contract NAS 5-26555. These observations are associated with program \#13432}}

\author[Kerzendorf, W.E.~et al.]{
W.E.~Kerzendorf,$^{\!1, 2}$\thanks{E-mail: wkerzendorf@gmail.com} %
Knox S.~Long,$^{\!3, 4}$ %
P.~Frank Winkler,$^{\!5}$ %
Tuan Do,$^{6}$
\\
$^{1}${European Southern Observatory, Karl-Schwarzschild-Stra{\ss}e 2, 85748 Garching bei M\"{u}nchen, Germany}\\
$^{2}${ESO Fellow}\\
$^{3}${Space Telescope Science Institute AURA, 3700 San Martin Drive, Baltimore, MD 21218, USA}\\
$^{4}${Eureka Scientific, Inc. 2452 Delmer Street, Suite 100, Oakland, CA 94602-3017, USA}\\
$^{5}${Department of Physics, Middlebury College, Middlebury, VT 05753, USA}\\
$^{6}${Physics and Astronomy Department, University of California, Los Angeles, USA}
}
\date{Accepted XXX. Received YYY; in original form ZZZ}

\pubyear{2018}

\begin{document}
\label{firstpage}
\pagerange{\pageref{firstpage} -- \pageref{LastPage}}
\maketitle

\begin{abstract}
If some or all Type Ia SNe arise from accretion onto a massive WD from a companion, then the companion will remain in some form after the SN explosion. Tycho-B is an unusual, relatively hot star along the line of sight to Tycho's SNR -- conclusively shown to be a Type Ia -- and has been suggested as such a companion. If the interior of Tycho's SNR contains unshocked Fe, and if Tycho-B is either within the SNR shell or in the background, then one might hope to see evidence of this in the UV spectrum.  Such is the case for SN\,1006, where spectra of the background Schweizer-Middleditch star, as well as two AGNs, show broad absorption lines of Fe II. To test this idea, we have used STIS on HST to obtain a UV spectrum of Tycho-B. The observed spectrum, however,  shows no evidence of Fe II absorption. Furthermore, a luminosity distance estimate using UV and optical spectra of Tycho-B suggests that the star is consistent with a foreground interloper. We conclude either that Tycho B is nearer than Tycho's SNR, or that all of the Fe in the interior of Tycho's SNR is more highly ionized.
\end{abstract}


\maketitle

\section{Introduction}

\sneia are of great interest to astronomy, not only as end-points of stellar evolution, but  also  as one of the most powerful cosmological distance probes. It is therefore unfortunate that the progenitor evolution is not understood. While there is consensus that the \snia phenomenon is powered by the thermonuclear explosion of a relatively massive ($>1\msun$) CO white dwarf, it is unclear how this thermonuclear run-away is triggered or how such a massive object is created. One possibility---known as the singly-degenerate (SD) scenario---is that  the white dwarf grows to a mass of 1.38\msun\ through accretion from a companion star and then self-ignites when the center reaches $\rho>10^9$~\si{\gram\per\cubic\cm}. While the massive white dwarf explodes, the companion star will survive in most cases \citep[e.g. ][]{2000ApJS..128..615M, 2008A&A...489..943P,2013ApJ...773...49P, 2013ApJ...765..150S}.  

The most popular alternative is the doubly degenerate (DD) scenario: the merger of two degenerate objects (white dwarfs or stellar cores),  leading to ignition \citep{1984ApJ...277..355W,1984ApJS...54..335I,2010ApJ...722L.157V, 2003ApJ...594L..93L,2011MNRAS.417.1466K}. Another possibility starts by a detonation wave running around the outer accreted helium layer of the white dwarf \citep[see e.g. ][]{2010A&A...514A..53F,2014ApJ...797...46S}. This sends shocks that coalesce at the center and raise the densities and temperatures resulting in a runaway thermonuclear detonation.

\begin{figure*}
	\includegraphics[width=\textwidth]{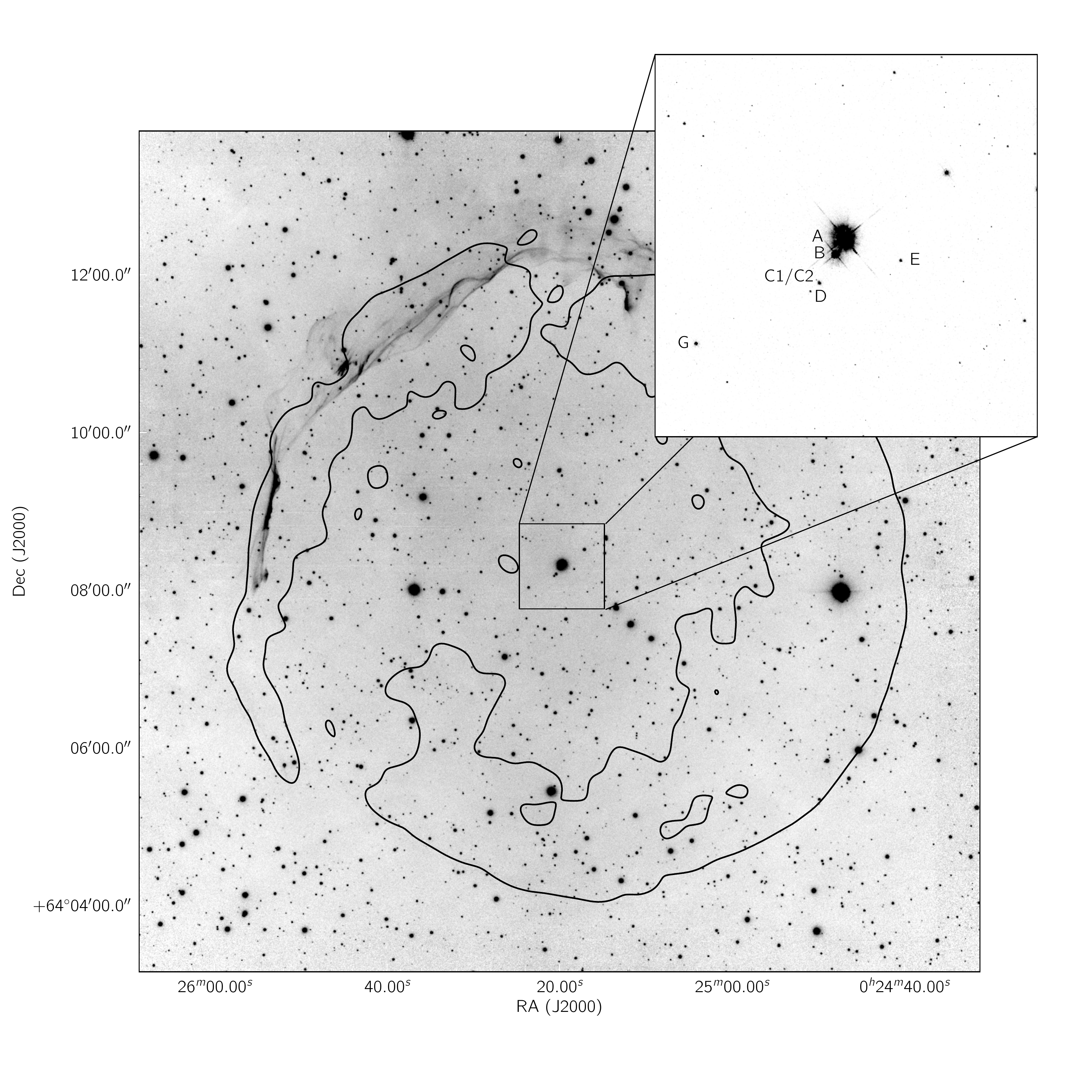}

    \caption{H$\alpha$ imagery from the 3.5m WIYN telescope at KPNO, obtained in September 2009 \citep{2015AAS...22514017P} of the remnant of \sn{1572}{} with contours from Chandra x-ray image (4.1 -- 6.1 keV). The detailed inset shows an {\em HST} WFPC2 image, using the F555W filter \citep{2004Natur.431.1069R}, of the center showing the candidate star Tycho-B and the other candidates near the center. }
    \label{fig:sn1572_overview}
\end{figure*}

The accretion (SD) scenario makes a clear prediction that would not occur in alternatives: the donor companion star survives the explosion. This has motivated a number of searches for such survivors in various \snia remnants. \citep[e.g.][]{2004ApJ...612..357R,2009ApJ...691....1G,2009ApJ...701.1665K,2014ApJ...782...27K, 2012Natur.481..164S}.  Here we focus on the efforts involving the remnant of Tycho's supernova (\sn{1572}). 

Tycho is a young SNR whose x-ray emission is dominated by emission from a reverse shock that is still propagating into ejecta from the explosion.
Inside the reverse shock, the remaining ejecta are expected to freely expanding, cold, and not highly ionized.  In the remnant of SN\,1006, also widely believed to have been a \snia (but without a light-echo confirmation), the cold ejecta have been observed spectroscopically through absorption from Fe (and Si)  using light from a background sub-dwarf B star \citep[now known as the SM star][]{1980ApJ...241.1039S} and two fainter AGN \citep{1983ApJ...269L...5W, 1993ApJ...416..247W, 2005ApJ...624..189W}, all of which provide ``core samples'' through the SN\,1006 shell.  

A number of searches for surviving companions in Tycho's SNR have been carried out, and several candidates have been suggested \citep[see][]{2004Natur.431.1069R, 2009ApJ...691....1G, 2009ApJ...701.1665K, 2013ApJ...774...99K, 2014MNRAS.439..354B}.

There are three promising  progenitor candidates: Tycho-G \citep[see][]{2004Natur.431.1069R}, Tycho-E \citep{2007PASJ...59..811I}, and Tycho-B \citep{2013ApJ...774...99K}. While Tycho-E and Tycho-G have some unusual characteristics compared to field stars \citep[see][respectively]{2004Natur.431.1069R, 2007PASJ...59..811I}, these sources can either be explained by normal stellar evolution \citep[Tycho-G;][]{2013ApJ...774...99K} or are too far from the remnant to be a plausible candidate  \citep[Tycho-E, at distance $\sim 10$~kpc;][]{2013ApJ...774...99K}.
 
The most unusual star near the center of the Tycho remnant (see Figure~\ref{fig:sn1572_overview})  is Tycho-B, first suggested as a candidate by \citet{2013ApJ...774...99K}. With an effective surface temperature $\teff\approx 10,000$ K, surface gravity $\logg \approx 4.0$, and metallicity $\textrm{[Fe/H]}\approx -1$, it is a young metal-poor star within the disk of the Galaxy and exhibits enhancements in carbon and oxygen. It also exhibits a relatively high rotational velocity ($\vrot \approx 170 \kms$) which is, however, not unusual for A-stars . \citet{2012arXiv1210.6050T} have suggested that Tycho-B, as surviving companion, might be explained if SN\,1572 resulted from a quadruple-star system.  
    
Tycho-B is bright enough in the UV that if it lies either within or beyond Tycho's SNR, and if Tycho's SNR still contains significant amounts of Fe$^+$, then one might hope to find broad absorption lines of Fe II in its spectrum, as is the case for the objects behind SN\,1006.  
To test this possibility, we have obtained UV spectra of Tycho-B with HST/STIS (GO 13432). Of the stars near the (projected) center of Tycho's SNR. Tycho-B is the only one bright enough in the UV to permit such an experiment, the results of which we report here.
We describe the observations and data reduction in Section~\ref{sec:obs_datared}, followed by presentation of our detailed analysis in Section~\ref{sec:analysis}. We then discuss our key results in Section~\ref{sec:discussion} and summarize our results in Section~\ref{sec:summary}.

\section{Observations and Data Reduction}
\label{sec:obs_datared}

We observed Tycho-B in a single two-orbit visit on 2014 September 30 using the STIS
 low-resolution grating (G230L) and the NUV-MAMA detector with the $52\arcsec\times0.5\arcsec$ slit, which provided wavelength coverage from 1570 to 3180 \AA\ with a resolution of about 3 \AA, appropriate for searches for broad absorption lines.  Photometry of earlier UV images (from HST Prop ID 6435) shows that Tycho-B has U-band (F336W) magnitude 16.78, bright enough that a relatively short HST/STIS observation should yield high enough signal-to-noise to  detect broad Fe~II absorption features if these are present.  
We eliminated most contamination from the  nearby ($2.7\arcsec$) and brighter (U = 15.96) star Tycho-A  by rotating the slit perpendicular to a line connecting  Tycho-A and Tycho-B. Two spectra were obtained with exposure times of 2630 and 3238 s. (The exposures were different in length because a portion of the first orbit was taken up with acquisition, using Tycho-A).  

\begin{figure}
	\includegraphics[width=0.5\textwidth]{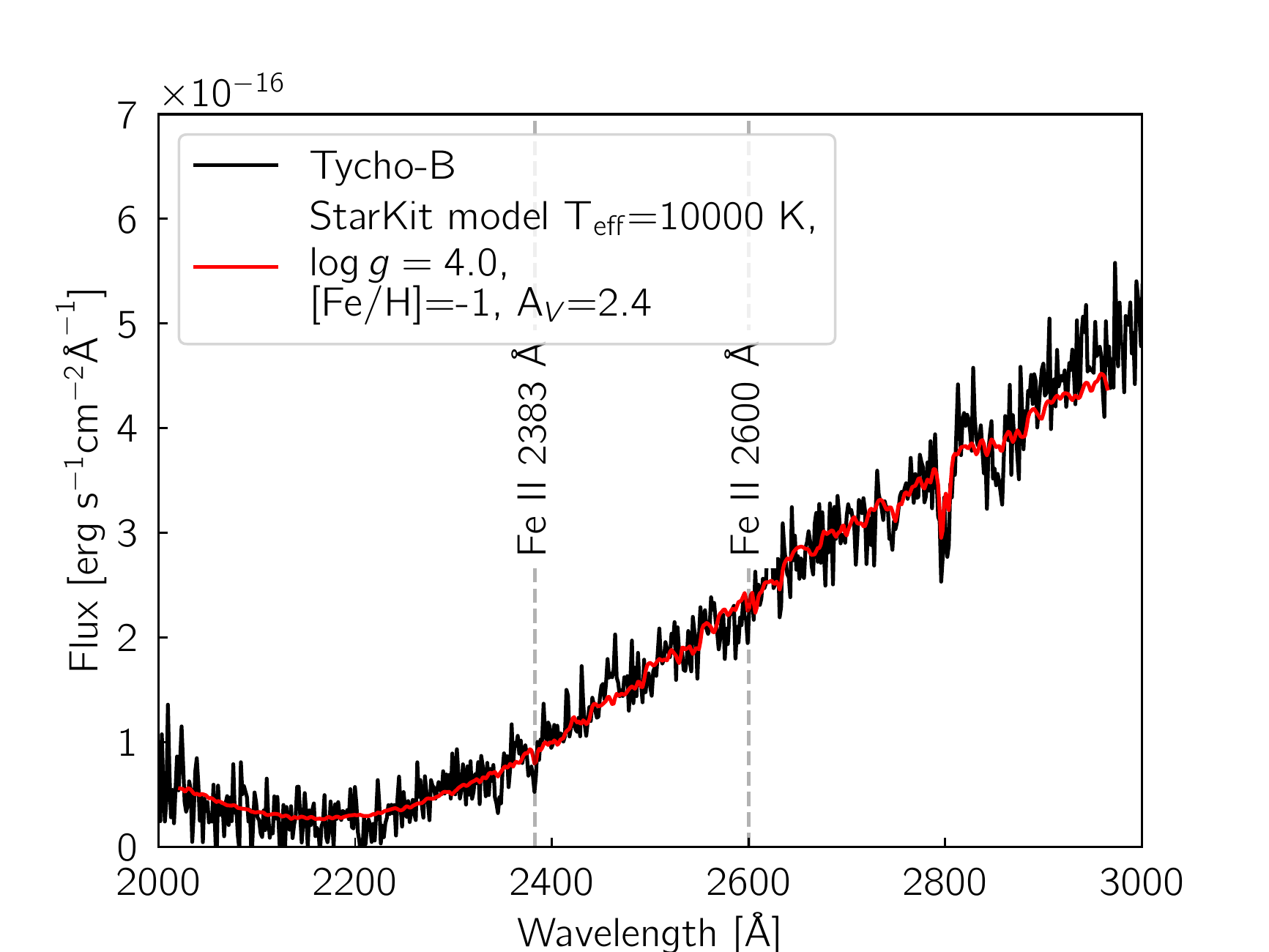}\\
	\includegraphics[width=0.5\textwidth]{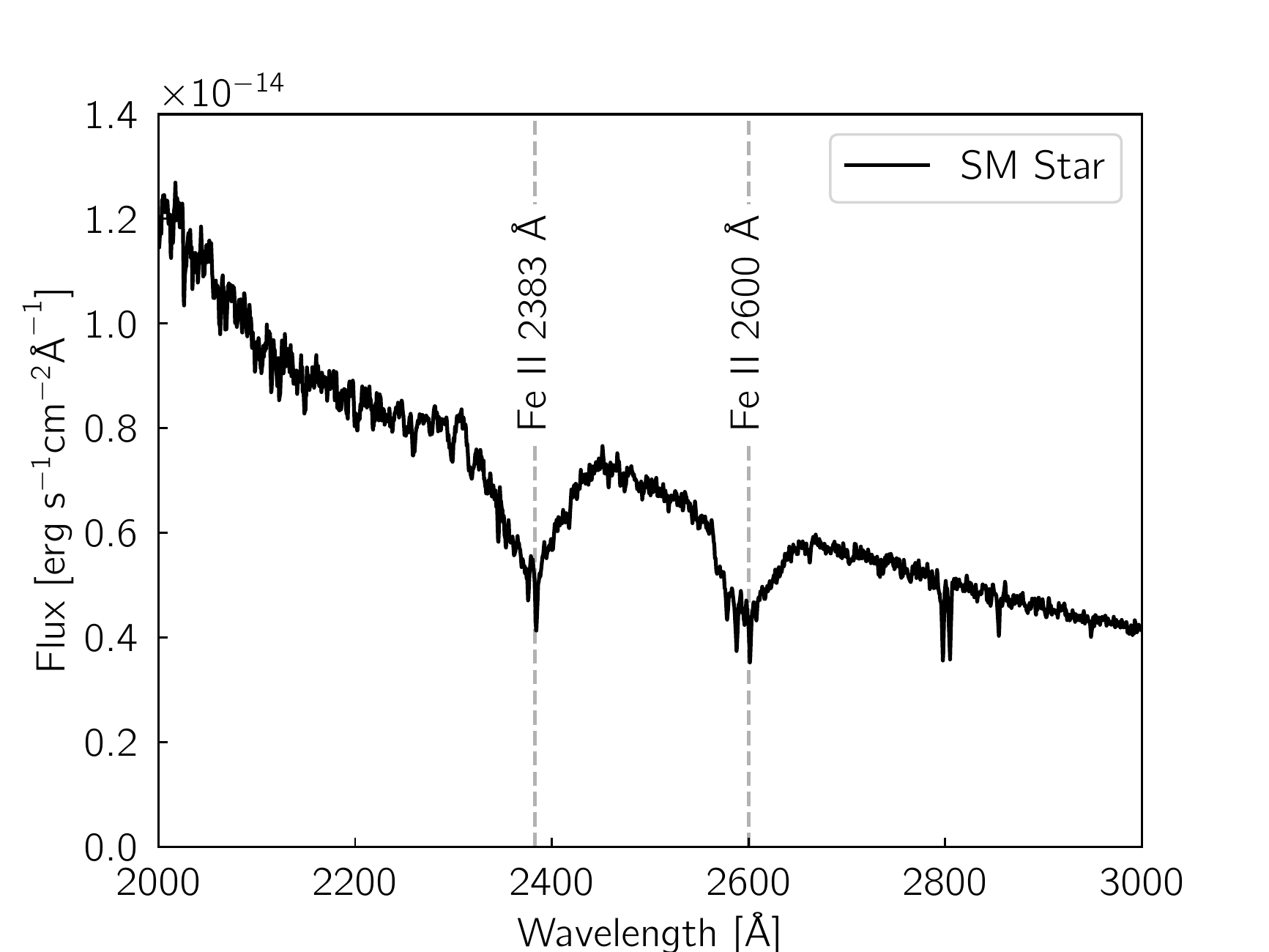}

    \caption{\textit{Upper Panel:} The average of the observed STIS spectra of Tycho-B overplotted with a stellar model given the parameters in \citet{2013ApJ...774...99K} \textit{Lower Panel:} A Faint Object Spectrograph (HST) spectrum of the Schweizer Middleditch star from \citep{1993ApJ...416..247W} showing the strong Fe absorption features from the foreground remnant SN\,1006. }
    \label{fig:simple_spec}
\end{figure}

The data used in our analysis were processed using version 3.4.1 of the STIS pipeline (in Sept. 2017).  
The average of the extracted x1d spectra, along with a model with the stellar parameters from \citet{2013ApJ...774...99K}, can be seen in Figure~\ref{fig:simple_spec}. For comparison we also show the UV spectrum of the \citet{1980ApJ...241.1039S} star that is behind the Type Ia remnant SN\,1006, which  shows prominent broad Fe~II absorption lines---features that are completely missing from Tycho-B's spectrum.  


\section{Analysis}
\label{sec:analysis}

In order to quantify our basic observational result that the spectrum does not contain absorption lines similar to those seen in SN\,1006. we have carried out a detailed Bayesian analysis of the spectrum, in an attempt to answer two key questions:
(1) How stringent are the limits on  possible broad Fe~II absorption?  (2) What is the allowed distance range for Tycho-B, and how does this compare with the distance of Tycho's SNR? 


\subsection{Models}
For modeling the spectrum, we generated synthetic spectra using the \gls{starkit} framework. Within \gls{starkit}, we used the \textsc{phoenix} grid of synthetic spectra  \citep{2013A&A...553A...6H} which spans $2300\:{\rm K}<\teff<12000\:{\rm K}$, $0.0<\logg<4.5$, and $-1.5<[{\rm M/H}]<1$. 
Starting from a model intrinsic spectrum, we then (a) convolve for a given \vrot (assuming a limb darkening of 0.6), (b) 
shift the spectrum for a given $\vrad$, (c) apply extinction parametrized with $A_V$ (assuming $R_V=3.1$) according to \citet{1989ApJ...345..245C},  (d) convolve with the appropriate instrumental profile (in our case $\Delta\lambda = 3\textrm{\AA}$), and (e) interpolate on the wavelength grid to match the observed data.  The flux of this synthetic model is then scaled to match the observed spectra of Tycho-B. This results in the stellar model  $M_\textrm{Tycho-B}(\teff, \logg, \feh, \vrot, \vrad, A_V) \mapsto F_\lambda(\lambda)$. 

We assume that the most prominent absorption features from the remnant would be the Fe~II lines at 2383\,\angstrom\ and 2600\,\angstrom, respectively.
We model the amplitude of these two features independently, but assume that both have the same velocity broadening profile. Our model sees Tycho-B being within the remnant (no red-shifted component). The spectrum does not show any obvious absorption features and thus a pure in-remnant model is enough to quantify how much absorption might be possible without being immediately visible. 
The model is then $M_\textrm{absorption}(A_{2383}, A_{2600}, \sigma, v_\textrm{remnant})\mapsto  \textrm{Transmission}(\lambda)$.


\subsection{Priors}

We choose the uncertainties given in \citet{2013ApJ...774...99K} as priors for our model of Tycho-B: $\teff=10,000^{+200}_{-400}$~K, [M/H] = $-1 \pm 0.4$, $\vrot=171^{+16}_{-33}$~\kms, $\vrad=-51\pm2$~\kms. We use the maps provided by \citet{2015ApJ...810...25G} to obtain the extinction between 1 -- 5~kpc and obtain a uniform prior for  $A_V=$2.2 -- 3.1. We assume a uniform prior for the amplitude for any absorption feature in our models of 0 -- 1. We require the velocity shift imposed by the remnant's expansion of 0 -- 6000$~\kms$ \citep[using the expansion velocities by][as a guide]{2017ApJ...840..112S}. We assume a uniform prior for the broadening of 500 -- 6000$~\kms$ \citep[using the uncertainties in the expansion velocities as a lower limit and the expansion velocities as an upper limit by ][]{2017ApJ...840..112S}.

\begin{figure}
	\includegraphics[width=0.5\textwidth]{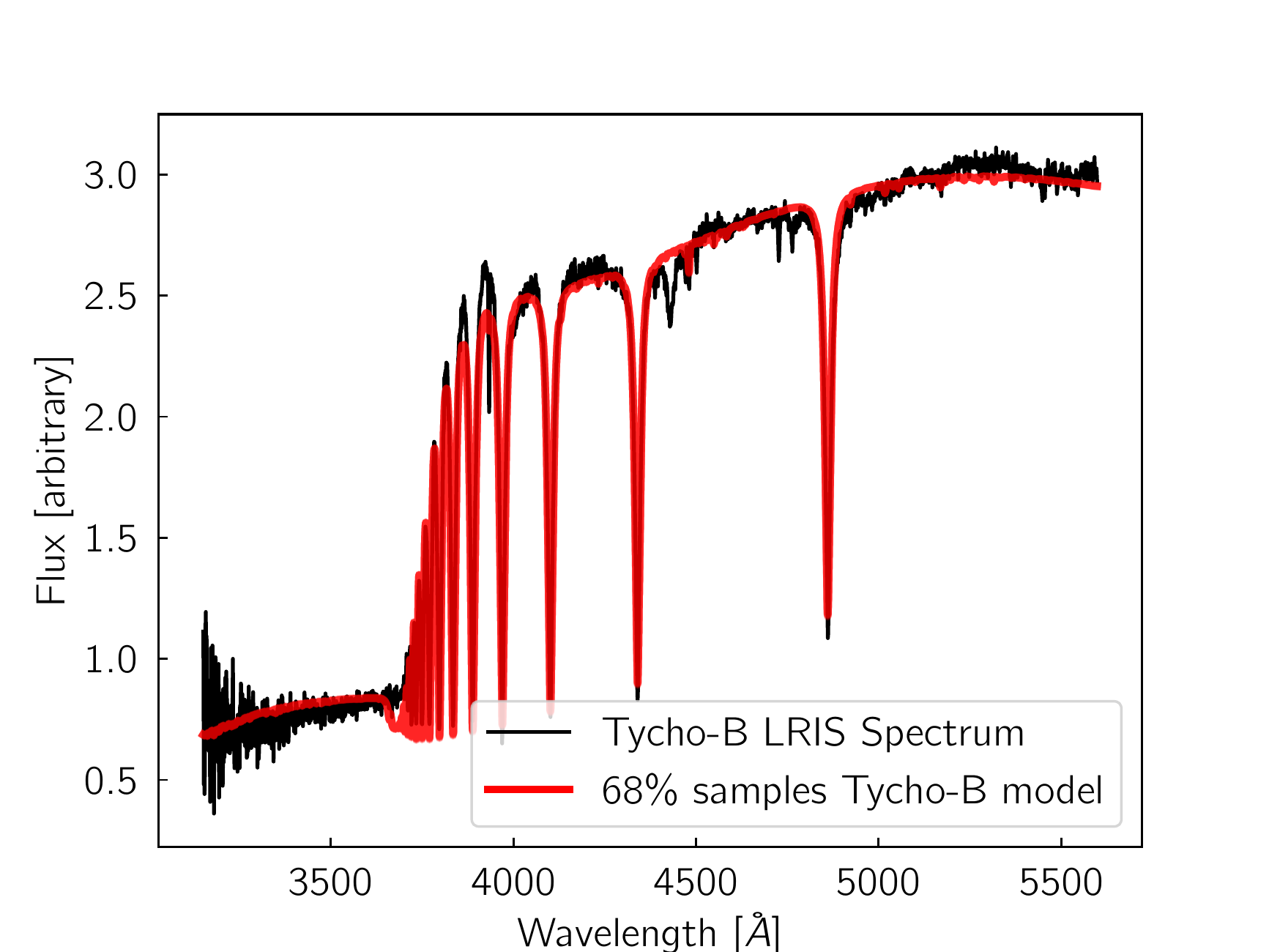}\\

    \caption{Tycho-B LRIS spectrum \citep{2013ApJ...774...99K} with a \textsc{StarKit} fit used to constrain \logg.}
    \label{fig:tychob_lris_fit}
\end{figure}

The determination of the surface gravity is of crucial importance for the distance estimate. The stellar features in the current STIS spectrum (mainly the Mg\textsc{ii} doublet at 2800\AA) is only very mildly sensitive to the surface gravity. \citet{2013ApJ...774...99K} provide an LRIS spectrum (3200\AA -- 5600\AA) covering the $\log{g}$ sensitive Balmer break region. This previous analysis -- using a predecessor of \textsc{StarKit} --  provided only a rough estimate for the uncertainty. We have redetermined the $\log{g}$ posterior probability using the setup described in this work  for the spectral fit (see Figure~\ref{fig:tychob_lris_fit}). We note that the large absorption feature missing in our models near $4400$\AA\ is a diffuse interstellar band. The fit gives an extremely tight 68\% quantile ($\log{g}=4.23 \pm 0.01$) which we will use as a $\log{g}$ prior for our Tycho-B fit. We note that the uncertainty is certainly underdetermined, and we take this into account in the discussion.

\subsection{Parameter inference}

We use the \gls{multinest} algorithm to infer the parameters (using the implementation available at \url{https://github.com/kbarbary/nestle}). The stopping criterion for such an algorithm is a comparison with an estimate of unaccounted evidence $Z_\textrm{est}$ when compared to the currently calculated evidence $Z_i$ for iteration $i$. We choose the default value of $\log({Z_i + Z_\textrm{est}}) - \log{Z_i} < 0.5$ for this criterion.

\begin{figure*}
	\includegraphics[width=\textwidth]{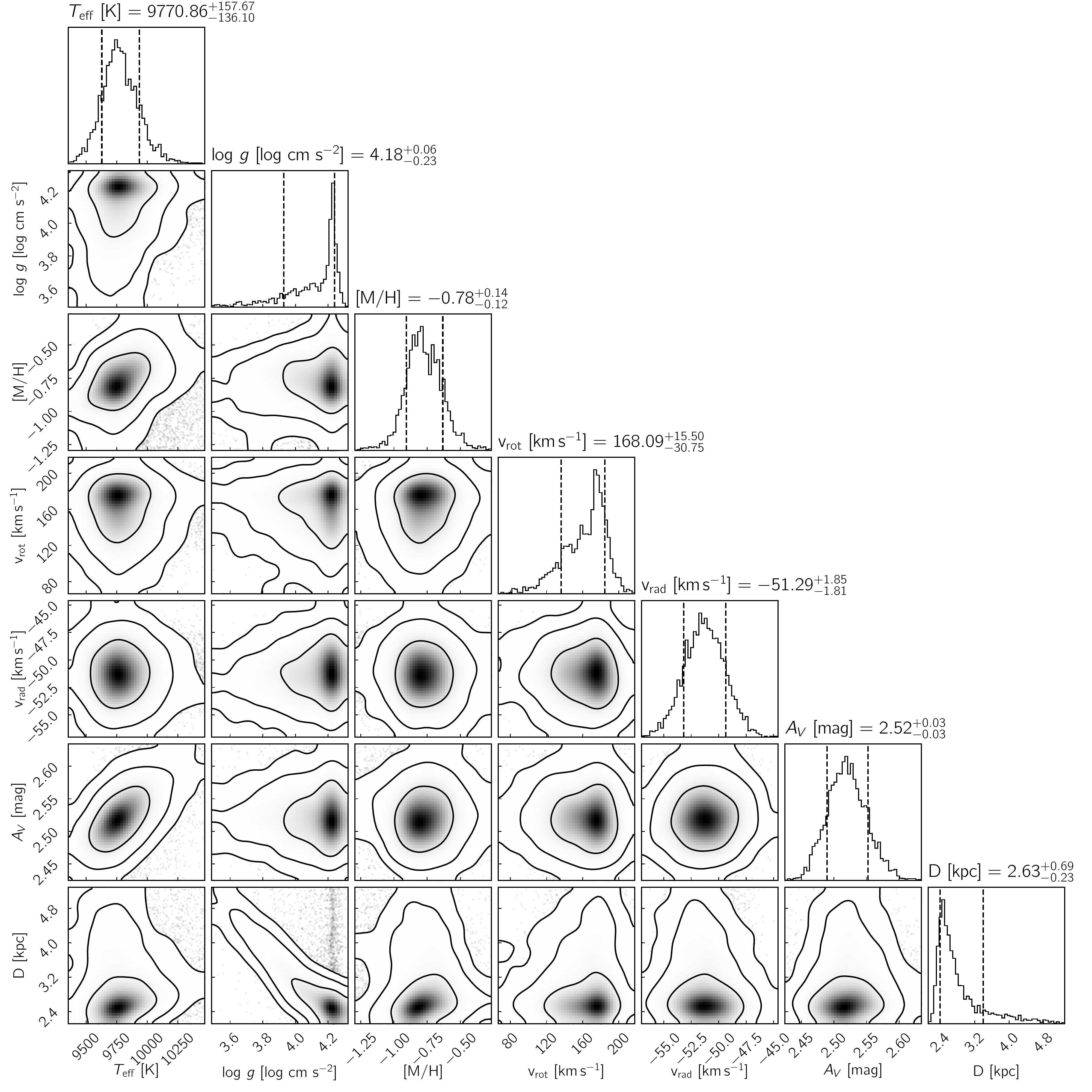}
    \caption{Stellar parameter estimation for Tycho-B using the presented \textsc{StarKit} model}
    \label{fig:tychob_corner_fg}
\end{figure*}

\begin{figure*}
	\includegraphics[width=\textwidth]{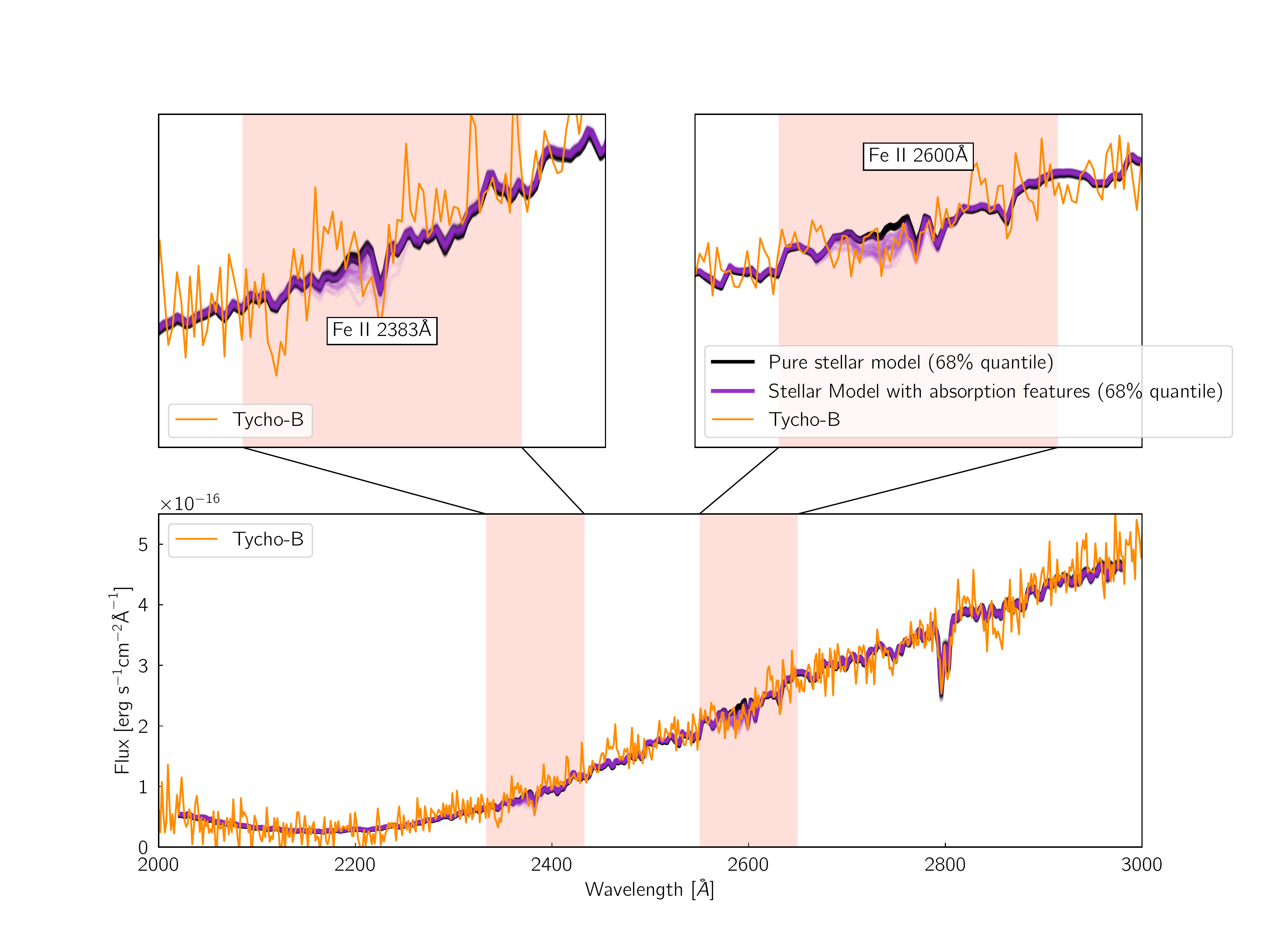}
    \caption{HST spectrum of Tycho-B in comparison with the modelling effort. We present the samples from the 68\% quantile (for both the models with and without remnant absorption). The cutouts center on the parts of the spectrum where the Fe\textsc{ii} absorption is expected to be. }
    \label{fig:tychob_spectrum}
\end{figure*}

We explore the parameter space for our models to match the data in three stages. We first explore the \citet{2013ApJ...774...99K} LRIS spectrum to get a prior for $\log{g}$ (see Figure~\ref{fig:tychob_lris_fit}). We then explore the stellar parameters (including the luminosity distance) using the flux calibrated STIS spectrum presented in this work and assuming the given priors (see Figure~\ref{fig:tychob_corner_fg}). Finally, we fit our model with the potential remnant absorption features and stellar parameters appropriate to Tycho-B. We show this posterior probability marginalized over the stellar parameters in Figure~\ref{fig:tychob_corner_in}. For both models (with and without remnant absorption features), we show a selection of fits from the 68\%-quantile in Figure~\ref{fig:tychob_spectrum}

\section{Discussion}
\label{sec:discussion}

\begin{figure*}
	\includegraphics[width=\textwidth]{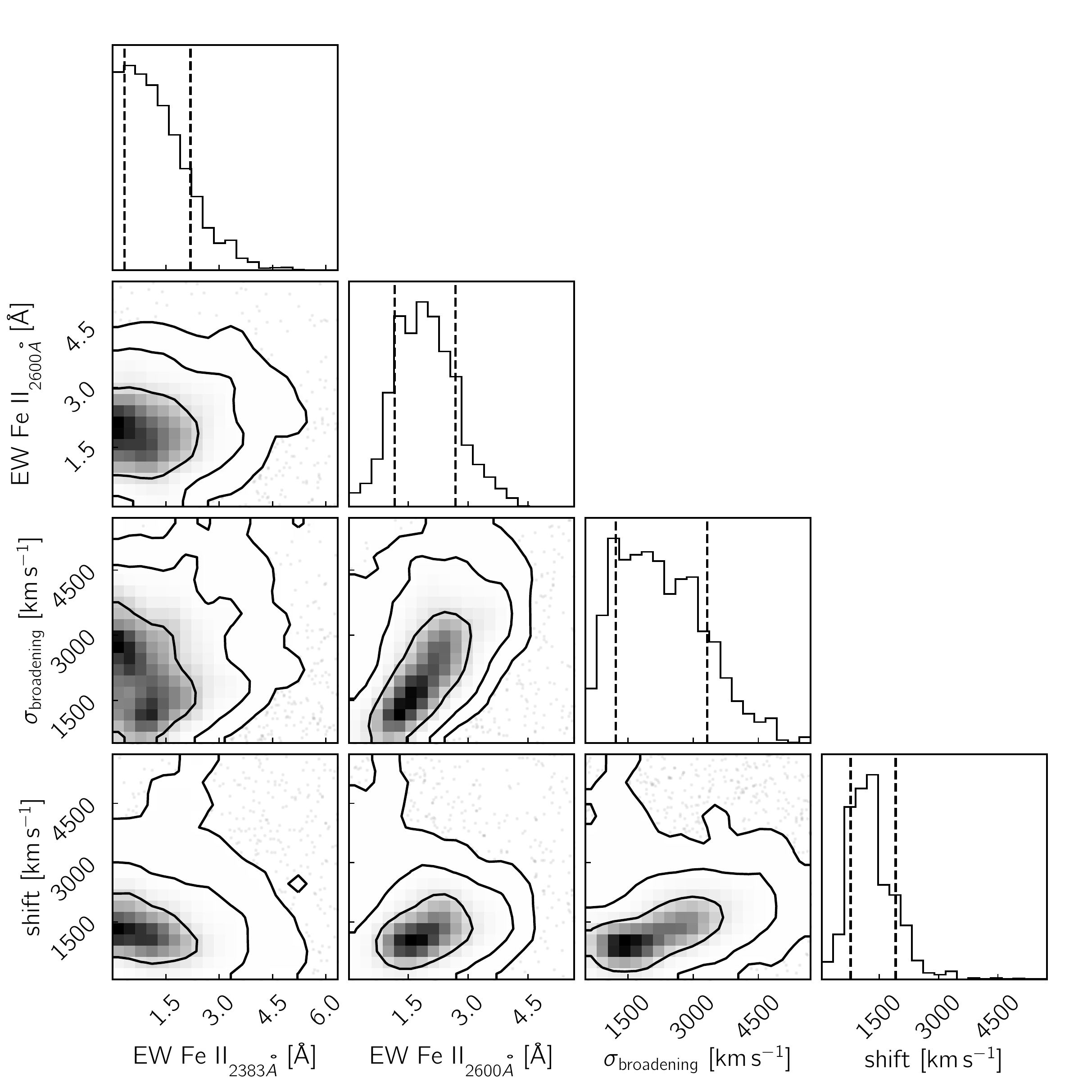}
    \caption{Parameter estimation for the absorption troughs for a model with the star within the remnant marginalized over the stellar parameters }
    \label{fig:tychob_corner_in}
\end{figure*}

In our detailed analysis, we try to quantify how much absorption is still possible in the Tycho-B spectrum given no absorption being immediately visible in the spectrum (see Figure~\ref{fig:simple_spec}). Figure~\ref{fig:tychob_corner_in} shows very low possible EW for the features (both of them likely upper limits) with $0.7_{-0.52}^{+1.03}$\AA\ and $2.19_{-0.93}^{+0.88}$\AA\ for the $\lambda 2382$~\AA\ and $\lambda2600$~\AA\ lines respectively. The EW for these lines measured in the SM star behind SN\,1006 is much larger:  15.4~\AA\ and 14.8~\AA\ for the $\lambda 2382$~\AA\ and $\lambda2600$~\AA\ lines respectively. Tycho-B  could be within or behind the remnant, but only if the column density of Fe~\textsc{ii} is extremely low.

One explanation might be that the density of cool iron is much lower than expected, resulting in no absorption features \citep[despite similar features in the \snia remnant SN\,1006;][]{2005ApJ...624..189W}. 
In this regard, it should be pointed out that our detailed understanding of Fe in SN\,1006 is not as precise as one might like.  The total amount of Fe II (and Fe III) in SN\,1006 is considerably less than expected for a \snia explosion, 0.2 -- 0.3\msun. \citet{2007MNRAS.381..771H} report a value of 0.044 \msun with a 3$\sigma$  limit of 0.16 \msun.  \citet{1988ApJ...327..178H} interpreted this as due to photon ionization from the reverse shock in SN\,1006. (They also predicted that the lines in Tycho's SNR should be similar to those in SN\,1006.)  Subsequently,  primarily as a result of observations obtained of Fe III lines in the FUV and an analysis of lines of Si II, \citet{2007MNRAS.381..771H} concluded that Fe is not very highly ionized in the interior of SN\,1006. 


To assess these possibilities, we next discuss independent distance measurements for both the Tycho SNR and Tycho-B.

\subsection{Distance to the Tycho SNR}
\label{sec:distance}
The distance to the Tycho SNR is itself quite uncertain.  Reconstructions of the light curve based on 16th century records from Tycho and others have long provided evidence that it was a Type Ia event, with apparent visual magnitude at maximum of $-4.0 \pm 0.3$ \citep{1945CMWCI.711....1B, 2004ApJ...612..357R}. 
\citet{2008Natur.456..617K}  observed the light-echo spectrum from SN\,1572 (over four centuries later) and showed conclusively that it had been a Type Ia event. Comparison with several template SN Ia spectra shows that it was a normal Type Ia, and correcting for extinction, estimated by \citet{2004Natur.431.1069R} to be $A_V = 1.86 \pm 0.12$, led \citet{2008Natur.456..617K} to a distance estimate of $3.8 ^{+1.5}_{-1.1}$ kpc. 

Radio  measurements of H~I absorption to Tycho can be used to estimate the distance kinematically through comparison with Galactic rotation curves, with the difficulty that Tycho is located in the outer Perseus arm, where a spiral shock causes a velocity reversal.  \citet{2011ApJ...729L..15T} reported the most comprehensive kinematic measurement and review others to arrive at a distance of 2.5 - 3.0 kpc.  

In a recent X-ray study, \citet{2010ApJ...725..894H} use ejecta radial velocities measured from {\em Suzaku} and proper motion measurements from {\em Chandra}, to obtain a distance estimate of $4.0 \pm 1.0$ kpc.  (They also provide a  review of measurements by other techniques.)  A similar but more detailed analysis based on {\em Chandra} data alone by \citet{2017ApJ...840..112S} arrived at essentially the same distance.  In both these studies, determining the distance relies on some assumptions about the geometry, since the proper motions are measured at the rim, while radial velocities are for interior knots.

Proper motions of the outer optical filaments, combined with the shock velocity inferred from the width of the broad Balmer lines that characterize them, lead to closer distances, 2.3 - 3.1 kpc \citep{1980ApJ...235..186C, 1991ApJ...375..652S,  2001ApJ...547..995G}.  While both the proper-motion and velocity-broadening measurements are done for the same outer filaments and are quite precise, the difficulty in this method comes in the shock models that are necessary to infer a shock velocity from the velocity broadening. These must account for energy lost to the acceleration of charged particles at the shock front--- still somewhat uncertain \citep[e.g.,][]{2010PASA...27...23H, 2013A&A...557A.142M}.

\citet{2012A&A...538A..81M} estimate the distance as $\sim 3.3$ kpc, based on their models to reproduce the extremely faint $\gamma$-ray emission from Tycho  detected from VERITAS \citep{2011ApJ...730L..20A} and Fermi-LAT \citep{2012ApJ...744L...2G}. \citet{2014ApJ...783...33S} have since developed a more sophisticated hydrodynamic model for the broadband spectrum of Tycho from radio through $\gamma$-rays to arrive at a similar distance estimate of $\sim 3.2$ kpc.

\begin{figure}
	\includegraphics[width=0.5\textwidth]{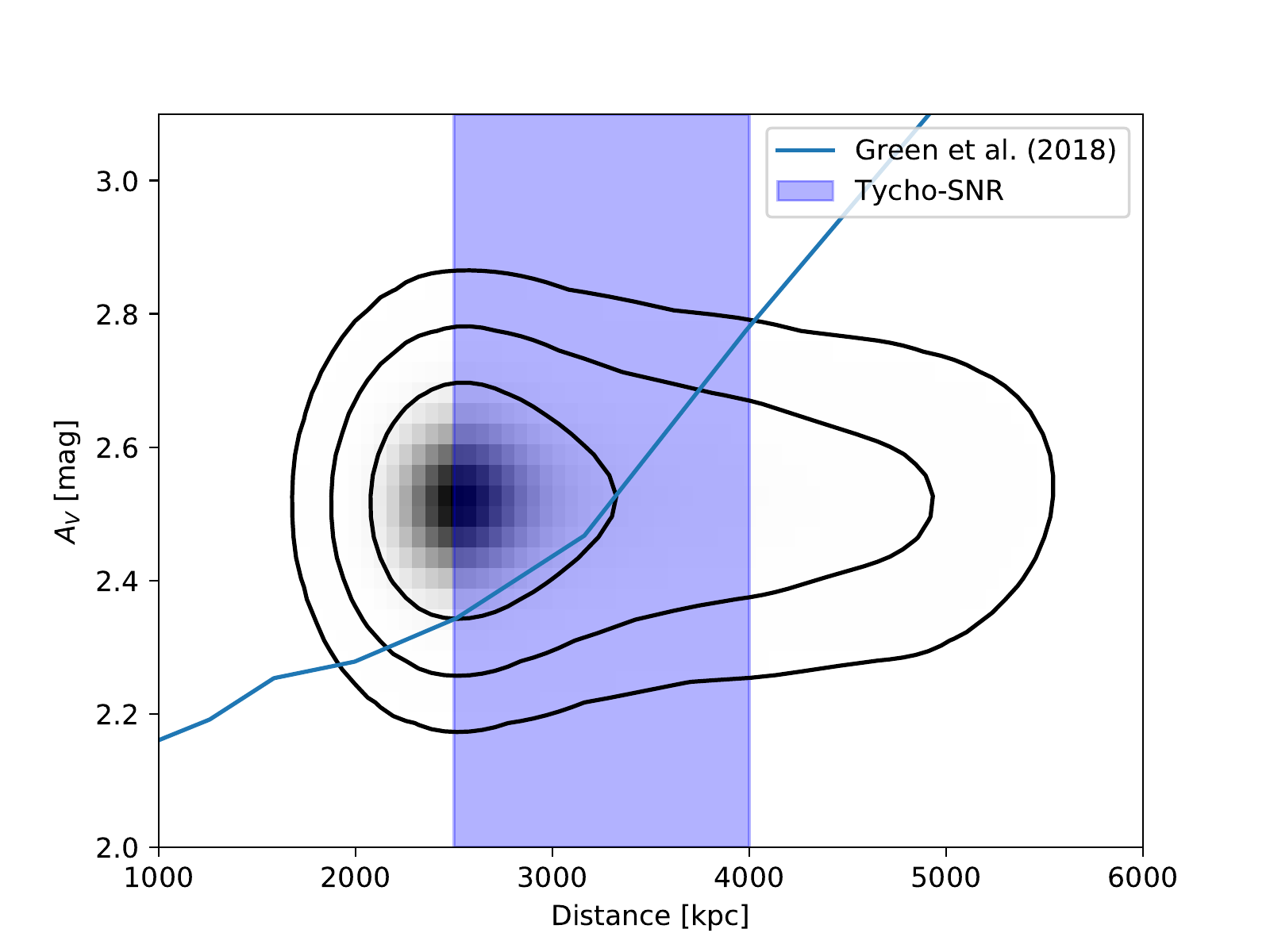}
    \caption{A comparison of distance and extinction inferred from the STIS data (contours mark the 68\%, 95\%, and 99.7\% quantiles) with distance and extinction relationship inferred by \citet{2018arXiv180103555G}. We have also marked a conservative distance estimate for Tycho's supernova remnant as discussed in the text.} 
    \label{fig:a_v_distance}
\end{figure}

A distance range of 2.5 - 4.0 kpc to Tycho, which we have adopted for comparison with Tycho-B in Figure~\ref{fig:a_v_distance} (discussed below) embraces virtually all of the recent estimates.  However, a more precise measure of the distance to Tycho would clearly be valuable.

\subsection{Distance to Tycho-B}

We have found the luminosity distance to Tycho-B by using the STIS UV spectrum in combination with the optical spectrum from LRIS \citep{2013ApJ...774...99K}.  For modeling the spectrum, we generated synthetic spectra using the \gls{starkit} framework, and then apply Bayesian statistics to obtain the allowed luminosity range, as detailed in Section~\ref{sec:analysis}.  Since the STIS spectrum is well flux-calibrated, we can then obtain a luminosity distance. 

The result is consistent with the stellar parameters presented in \citet{2013ApJ...774...99K}. The distance determination of $2.63 ^{+0.69}_{-0.23}$~kpc is consistent with Tycho-B being in front of the remnant. However, we note that this distance estimate is very sensitive to the $\log{g}$ measurement, which we believe carries a larger systematic uncertainty that is not included in the determination. Thus we have tried to independently check our results. 

One possibility is to use the extinction-distance relation determined by \citet{2018arXiv180103555G} to obtain an independent distance estimate given our inferred $A_V$ measurement. They use stars as light sources to measure their foreground dust column and infer the distance and brightness of the star by employing probabilistic models. Figure~\ref{fig:a_v_distance} shows that for our inferred $A_V$ we obtain a distance similar to the luminosity distance. We believe the difference to arise from the various systematic uncertainties that are not directly included in the fit (e.g. the absolute luminosity of the models, the $\log{g}$ sensitivity of the models, etc.). We have also marked the distance uncertainty for Tycho's supernova remnant.

\section{Summary}
\label{sec:summary}

Tycho-B is one of several stars that have been proposed as a progenitor companion for the object that produced Tycho's SN.  In order to shed light on this issue, we have obtained UV spectra of Tycho-B with HST/STIS.  We hoped to use these spectra to identify absorption near 2383 and 2600 \AA, as is seen in UV spectra of objects behind SN\,1006, that might---depending on the shape of the absorption lines---indicate that Tycho B was within or behind the Tycho SNR.  Our principal findings are as follows:

\begin{itemize}
\item
There is no evidence of broad absorption near 2383, 2600 \AA.  Our upper limit on the equivalent width of these  absorption features is a few Angstrom (on a 3$\sigma$ level; see Figure~\ref{fig:tychob_corner_in}), compared to the $\approx 15$\AA\ EW for the broad lines seen in spectra of the Schweizer-Middleditch star which lies behind SN\,1006.  This implies either that Tycho-B is in front of Tycho's SNR, or that Fe in the interior of Tycho's SNR exhibits a higher ionization state.
\item
The spectrum of Tycho B is consistent with that of a 10,000 K main sequence star, and the detailed stellar parameters are consistent with the analysis in \citet{2013ApJ...774...99K}.  Both the luminosity distance and the inferred A$_V$ imply a distance of $\approx 2.6$~kpc with large uncertainties. 
While we cannot rule out the possibility that Tycho B is behind or within Tycho's SNR (and that interior Fe is highly ionized), our analysis favors its being a foreground object.
\end{itemize}

Overall, our conclusion is that Tycho-B is unlikely to be the progenitor companion of the object that produced Tycho's SNR. Tycho-B has been measured by Gaia (Gaia DR2 431160569875463936) and shows a distance of 1.9 -- 2.2\,kpc and thus is consistent with our conclusion that Tycho-B is a foreground star 	

The apparent absence  of plausible companions suggests that SN\,1572 was not produced by the classical accretion scenario.
This conclusion is similar to the one reached by \citet{2017NatAs...1E.263W} using a different argument.

\section{Acknowledgements}

W.~E.~Kerzendorf was supported by an ESO Fellowship and the Excellence Cluster Universe, Technische Universit\"at M\"unchen, Boltzmannstrasse 2, D-85748 Garching, Germany.  Support for program \#13432 was provided to K.~S.~Long  and P.~F.~Winkler by NASA through grant HST-GO-13432 from the Space Telescope Science Institute, which is operated by the Association of Universities for Research in Astronomy, Inc., under NASA contract NAS 5-26555.  P.~F.~Winkler acknowledges additional support from the NSF through grant AST-1714281. We would like to acknowledge the detailed discussions and input from Marten van Kerkwijk and Kim Venn during the preparation of the HST proposal that led to the observations reported here. We would like to acknowledge the helpful comments by our referee.

\bibliographystyle{mnras}
\bibliography{wekerzendorf}

\end{document}